\documentclass[12pt]{article}

\usepackage{amssymb,dsfont}
 \usepackage{amsfonts}
\usepackage{amssymb}
\usepackage{amsmath}
\usepackage{graphicx,psfrag, epsf}
\usepackage{enumerate}
\usepackage{natbib}
\usepackage{subcaption}
\usepackage{bbm,rotating}
\usepackage{dsfont}
\newcommand{\blind}{0}
\usepackage{cleveref}
\usepackage{mathptmx}

\newcommand{\E}{\mathbb E}

\usepackage{./StefanTexCommandsV2}

\graphicspath{{./pic/}}

\begin{document}

\newcommand{\SPACEBIG}{1.45}
\newcommand{\SPACESMALL}{1}

\def\spacingset#1{\renewcommand{\baselinestretch}%
{#1}\scriptsize\normalsize} \spacingset{1.5}

%%%%%%%%%%%%%%%%%%%%%%%%%%%%%%%%%%%%%%%%%%%%%%%%%%%%%%%%%%%%%%%%%%%%%%%%%%%%%%

\if0\blind
{
  \title{ Empirical Bayes approaches to PageRank type algorithms for rating scientific journals
%Empirical Bayes approaches to PageRank type algorithms for ranking scientific journals.
%Empirical Bayes of Google PageRank to the rescue of Bibliometrics
%Empirical Bayes interpretation of PageRank index in Bibliometry/ Google PageRank to the rescue of Bibliometrics
}
\author{Jean-Louis Foulley$^1$, Gilles Celeux$^2$ \& Julie Josse$^3$\\
1. Institut montpellierain Alexander Grothendieck, foulleyjl@gmail.com \\
2. INRIA,  gilles.celeux@inria.fr \\
3. Polytechnique, CMAP,  julie.josse@polytechnique.edu \\
}
  \maketitle
} \fi

\if1\blind
{
  \bigskip
  \bigskip
  \bigskip
  \begin{center}
    {\LARGE CA Poisso,}
\end{center}
  \medskip
} \fi

%\bigskip
%\begin{abstract}

%\end{abstract}

%\noindent%
%{\it Keywords:}  

\spacingset{\SPACEBIG}

\begin{abstract}

Following criticisms of the
journal Impact Factor, new journal influence scores have been developed such as the Eigenfactor or the Prestige Scimago Journal Rank. 
They are based on  PageRank type algorithms on the cross-citations transition matrix of the citing-cited network.
The PageRank algorithm performs a smoothing of the transition matrix combining a
random walk on the data network and a teleportation to all possible nodes with fixed probabilities. We reinterpret this smoothing matrix as the mean of a posterior distribution of a Dirichlet-multinomial model
in an empirical Bayes perspective. 
We suggest a simple yet efficient way to make a clear distinction between structural  and sampling zeros. This allows us to contrast cases with self-citations included or excluded to avoid overvalued journal bias.
We estimate the model parameters by maximizing the marginal likelihood with a Majorize-Minimize algorithm. The procedure ends up with a score similar to the PageRank ones but with a damping factor depending on each journal. The procedures are illustrated with an example about cross-citations among 47 statistical journals studied by \citet{CIS-448159}.  
\end{abstract}

\section{Introduction}

Assessing and ranking journals using influence indicators is an old practice \citep{Gross385}  which has grown with the introduction of the impact factor (IF) \citep{Archambault2009}.  The IF measures the reputation of a journal by the average number of annual citations it receives per article published in the last two years  \citep{Garfield471}. The systematic publication of IF by Clarivate Analytics (ex Thomson-Reuters, ex Institute of Science Information) in journal of citation reports (JCR) greatly impacts all sectors of scientific life and policies. The hierarchy created between journals generates intense competition among them, researchers and institutions. However, the IF is highly criticized for both technical \citep{Vauclay12} and ethical reasons \citep{Zitt12}. 
It does not take into account the critical (positive, neutral or negative) assessment of citations; it strongly depends on the disciplinary field; the citation window is too narrow (2 years); the asymmetric distribution of the number of citations of an article is poorly taken into account by the average; the self-citations may have a negative influence as well as the equal weight attributed to each quotation whatever its origin. Various alternatives have been proposed to deal with these issues: lengthening of the citation window, standardization by disciplinary field, etc. \citep{Zitt}.
%The most popular citation metric, the
%impact factor (Garfield 1972; Science) is
%controversial and widely abused.
%Citation behaviour varies greatly
%across disciplines, making inter-field
%comparisons difficult.
%More recent approaches based
%on the PageRank algorithm
%improve upon the impact
%factor by taking the source of
%citations into account.
%However, none of these
%methods includes any way of
%quantifying uncertainty,so are
%they really statistical?

Recent approaches taken into account the importance of citing sources in order to improve upon the IF have been suggested such as methods based on group lasso \citep{CIS-448159}, stochastic bloc models, clustering with modularity classes \citep{stigler1994, Arbel} or methods using symmetric row-column (RC) models \citep{Good85, Grah}. 

Alternative solutions include scores derived from the Google PageRank (PR) algorithm \citep{Walt} such as prestige scimago journal rank (PSJR) from Scimago Lab \citep{Gonz} which is released by Elsevier 
 and the  Eigenfactor (EIFA) from Eigenfactor$^{\mbox{TM}}$ Metrics \citep{Berg, West} which is released by Clarivate Analytics. They are attractive because of their simplicity and ease of computation.
However, such scoring procedures are not embedded in a probabilistic model framework  which  can be helpful to make the underlying assumptions explicit and their derivation mathematically rigorous. One  aim of this article is thus to suggest underlying models for these scores to better understand what is the rationale of the different operations applied.
We focus on EIFA which has the particularity of excluding self-citations in order to compensate for the biases in the incentive policies of certain journals and the harmful effects of a form of intellectual inbreeding.

% and to allow for assessment of the uncertainty in the ranking. 

The outline of the paper is as follows. In Section 2, we detail the different steps to obtain the EIFA score.
In Section 3, we show how this construction can be reinterpreted using a more formal empirical Bayesian perspective with a specific Dirichlet-multinomial  model. We derive a Majorize-Minimize algorithm for the inference.
We highlight the potential of the new method to appropriately handle structural zeros and distinguish them from sampling zeros, which is  always a key issue in statistical inference.
This ensures dealing with self-citations (inclusion or exclusion) and we show that our method results in a more flexible PageRank score.
In Section 4, we compare and contrast our proposition with PR and EIFA methods on an example to rank statistical journals.
A discussion section ends the paper. 
Our criteria are implemented as R functions and the code to reproduce all the results is available on Github \citep{Josse2017}. 

\section{PageRank influence scores}

Let $C \in \mathbb{N}^{N \times N}$ be a square matrix of the cross-quotations between $N$ journals of the same disciplinary field with citing (issuing references) in rows and cited (receiving citations) in columns.  More precisely,  $c_{ij}$ corresponds to the number of times journal $i$, in a given year, quotes articles published by the journal $j$ over a previous period of time (usually  2, 3 or 5 years). 
From this matrix, it is possible to define a weighted oriented network citing$\rightarrow$cited with the transition probability matrix $P$, with elements $p_{ij}= (\frac{c_{ij}}{c_{i+}})_{(i=1, \ldots, N), (j=1, \ldots, N)}$ where $c_{i+}= \sum_{j=1}^N c_{ij}$.
PageRank (PR) produces a smoothing of $P$ by using a convex combination of $P$ and a so-called "teleportation" matrix  $\frac{1}{N}\mathbf{1}_N \mathbf{1}_N^{\top}$ with $ \mathbf{1}_N$ a vector of 1s of size $N$ 
\begin{equation}  \label{eq:pagerank}
 G = \alpha P+ (1-\alpha)\frac{1}{N}\mathbf{1}_N \mathbf{1}_N^{\top},
\end{equation}
$\alpha$ being a scalar $\in (0,1)$.
%with  $\pi=\frac{1}{N}\mathbf{1}_N $. 
Thus, for all $i=1,\ldots,N$ and $j=1,\ldots,N$ 
\begin{equation}  \label{eq:pagerankgt}
 G_{ij}= \alpha P_{ij}+ (1-\alpha)\frac{1}{N}.
\end{equation}

%If some journals receive quotations but do not cite any other journal of the $(N -1)$ network, the %corresponding null row is substituted by the vector $\pi$ in $P$.  
%The teleportation  avoids of absorbing states and in particular disconnected components.
Equation \eqref{eq:pagerank} guarantees the graph associated to $G$ to be strongly connected and, thus, it defines the transition matrix of  of a discrete-time, irreducible and aperiodic Markov chain between the $N$  journals. 
%Note that in the case of zero rows in $P$,  corresponding to dangling nodes, \textit{i.e.} nodes without any outgoing citations, they are replaced by the vector $\pi$  to make $P$ a stochastic matrix \citep{Langville}. 
%Equation \eqref{eq:pagerank} can be illustrated with the famous "random surfer" of Google which starting from $i$, moves with probability $\alpha$ in the data network according to a random walk with transition probability $p_{ij}$, and  is randomly teleported, with a probability $(1-\alpha)$ to any other node with a transition probability $\pi_j$ independent of the node $i$.  
The parameter $\alpha$ is called the damping factor and is usually set to $0.85$. We will discuss this choice in Section 3.1.

%Irreductible (on arrive toujours à aller à un noeud en un nombre fini d'étapes)

%Do a random walk around the graph:
%1 Select a random journal and open it
%2 Select a random reference
%3 Open the cited journal
%4 Repeat 2–3, forever
%PageRank is the proportion of time spent
%reading each journal, i.e. the stationary
%distribution of an ergodic Markov chain.
%This was later adapted by Google in the 1990s
%for their search engine, with an added "damping
%factor" to allow teleportation around the graph.

The algorithm PR, as defined in the original article \citep{Brin:1998:ALH:297810.297827}, is a recursive algorithm which ranks the nodes according to the following score at iteration $\ell$:
\begin{eqnarray} \label{eq:pagerankpuissance}
r_j^{(\ell+1)}= \sum_{i=1}^{N} g_{ij} r_i^{(\ell)} \mbox{ , } j=1, \ldots, N.
\end{eqnarray}
This means that, in the calculation of the score for the journal $j$, the contribution of the citing journal $i$ is equal to its frequency $g_{ij}$ weighted by the proper current score $r_i$ of $i$. Thus, a quotation from a leading journal such as JRRS-B or JASA does not have the same weight as a quotation from another journal.
Since the influence scores are unknown, they are iteratively estimated according to the so-called power method. As a matter of fact, the limit value of \eqref{eq:pagerankpuissance} corresponds to the stationary distribution of the Markov chain with transition matrix $G$ and this stationary distribution is  independent of the initial state.
The solution is given by the eigenvector $r$  with unit norm 
%$(r^{\top} \mathbf{1}_N= 1)$ 
of $G$ associated with the greatest eigenvalue. Since 
the chain is irreducible and aperiodic, this eigenvalue is unique and equal to one \citep{Fouss}.
In practice, it means that  knowing $G$, it is possible to establish a ranking of the journals based on their importance.
%Using the Google's random surfing representation, the importance of nodes corresponds to the limit visit frequency of these nodes during a very long walk on the network for a large number of surfers whatever their initial position. \\

Other scores such as EIFA consider an  alternative  teleportation giving to Journal $i$ a weight proportional to $a_i$ the number of article published by $i$ in the time window considered. In fact, EIFA score has other differeneces with PR, in particular it excludes self-citations, but this point will be considered further later in this paper. 
Thus the resulting transition matrix has the general term 
\begin{equation}  \label{eq:eifagt}
 G_{ij}= \alpha P_{ij}+ (1-\alpha)\frac{a_i}{a_+},
\end{equation}
with $a_+=\sum_{i=1}^N a_i$. The scores (\ref{eq:pagerankgt}) and (\ref{eq:eifagt}) can be derived from a standard Dirichlet-multinomial  Bayesian model (see, for example, \citet{dirichletrank-solving-the-zero-one-gap-problem-of-pagerank}). This model is as follows.
 
\begin{enumerate}
\item The $(C_i=(C_{ij},j=1,\ldots,N); i=1,\ldots,N)$ are assumed to be independent and to follow a multinomial distribution $\mathcal{M}(n_i,  \theta_i)$, with $\theta_i=(\theta_i^1,\ldots,\theta_i^N)$.
 \item  The parameters $(\theta_i, i=1,\ldots,N)$ are assumed to be independent and to follow a Dirichlet prior distribution
$(\theta_i  | \gamma_{1}, \ldots,\gamma_N \sim \mathcal{D}(\gamma_1, \ldots,  \gamma_{N}); i=1,\ldots,N)$.
\end{enumerate}
Due to the conjugacy property of these distributions, the posterior distributions of the $\theta_i$ are Dirichlet distributions:
\begin{eqnarray}
\label{eq:diricpost}
\theta_i  | \gamma_{1}, \ldots,\gamma_{N}, C_i \sim \mathcal{D}(\gamma_{1}+c_{i1},\ldots,\gamma_{N}+c_{iN}), \mbox{for }i=1,\ldots,N.
\end{eqnarray}
Thus the posterior expectations of the $\theta_{ij}$ are
\begin{equation}
\label{eq:diricpostexptsim}
\E_{\mbox{post}}({\theta_{ij}})= \frac{c_{ij} + \gamma_j}{\sum_{j =1}^N c_{ij} + \sum_{j=1}^N \gamma_j}, \mbox{ for }  i= 1, \ldots, N; j=1,\ldots,N.
\end{equation}
Denoting $K=\sum_{j=1}^N \gamma_{j}$ and
$\alpha_i = \frac{n_i}{n_i + K}$, simple algebra leads to
\begin{equation} \label{eq:clg}
\E_{\mbox{post}}(\theta_{ij})= \alpha_i p_{ij}+ (1-\alpha_i)\frac{\gamma_j}{K}.
\end{equation} 
The shrinkage coefficient $(1 - \alpha_i)$  varies between 0 and 1. It is large when $n_i$ is small and $K$ is large.  When the data is not self-sufficient, \textit{i.e.} $K$ is large in comparison to $n_i$, there is a need to borrow information from other journals. The limiting case of  $\alpha_i = 0$ corresponds to a dangling node. In such a situation, the posterior expectations $\E_{\mbox{post}}(\theta_{ij})$ will be equal to the prior probabilities $\frac{\gamma_j}{K}$ for $j=1,\ldots,N$. 

If the hyperparameters are chosen to be equal for all the journals : $\gamma_j=\gamma$ for $j=1,\ldots,N$, we have $K=N\gamma$ and 
the equation (\ref{eq:clg}) becomes
\begin{equation} \label{eq:clgs}
\E_{\mbox{post}}(\theta_{ij})= \alpha_i p_{ij}+ (1-\alpha_i)\frac{1}{N}
\end{equation} 
and looks like equation (\ref{eq:pagerankgt}).

\section{A Bayesian Dirichlet-multinomial model without diagonal} \label{section:ebay}

Let $\underline C_i^{\top}= (c_{ij})$ for $j \neq i, i = 1, \ldots, N$ be the row $i$ of $C$ without its diagonal elements so that $\underline C$ is of dimension $(N \times N -1)$ and can be written $\underline C^{\top} = [\underline C_1, \ldots, \underline C_N]$.  
%We suggest an approach which has the property of handling in an integrated way both structural zeros (which correspond to self-citations) and sampling zeros. 
%This is a strong point of the method and has an impact for the ranking. 
%\subsection{Multinomial - Dirichlet}
We consider the following hierarchical probabilistic model. 
\begin{enumerate}
\item Multinomial sampling of the elements of  $\underline C_i^{\top}$
\begin{eqnarray}
\label{eq:multinom}
  \underline C_i^{\top}  | \underline \theta_i^{\top} \sim \mathcal{M}(n_i, \underline \theta_i^{\top})
\end{eqnarray}
 of parameters $n_i = \sum_{j \neq i} c_{ij}$
and probability vector elements $\underline \theta_i = (\theta_{i1}, \ldots, \theta_{i, i-1}, \theta_{i,i+1}, \ldots, \theta_{i,N})^{\top}$. 
The random vectors $ \underline C_i$ are assumed independent.
\item  Dirichlet prior distributions for the parameters of the distributions \eqref{eq:multinom}.  The prior distributions are also assumed independent:
\begin{eqnarray}
\label{eq:diric}
 \underline \theta_i^{\top}  | \gamma_{\backslash i} \sim \mathcal{D}(\gamma_{\backslash  i}^{\top})
\end{eqnarray}
where $\gamma_{ \backslash i}^{\top}= (\gamma_1, \gamma_2, \ldots, \gamma_{i-1}, \gamma_{i+1}, \ldots, \gamma_{N})$. 
\end{enumerate}
Due to the conjugacy property of these distributions, the posterior distributions are Dirichlet distributions:
\begin{eqnarray}
\label{eq:diricpost}
 \underline \theta_i^{\top}   | \gamma_{\backslash i}, \underline C_i^{\top}  \sim \mathcal{D}(\underline C_i^{\top} +\gamma_{\backslash i}^{\top}), ~ i=1,\ldots,N.
\end{eqnarray}
We easily get the expression of the posterior expectations:

\begin{eqnarray}
\label{eq:diricpostexpt}
\E_{\mbox{post}}(\underline {\theta_{ij}})= \frac{c_{ij} + \gamma_j}{\sum_{j \neq i} c_{ij} + \sum_{j \neq i} \gamma_j}, \mbox{ for } j \neq i, i= 1, \ldots, N.
\end{eqnarray}
Denoting
 \begin{eqnarray}
\label{eq:shrink}
\alpha_i = \frac{n_i}{n_i + K_{\backslash i}}, 
\end{eqnarray}
with $K_{\backslash i}= K - \gamma_i$ and $K=\sum_{i=1}^{N}\gamma_i$, formulae \eqref{eq:diricpostexpt} can be written as  $\E_{\mbox{post}}(\underline {\theta_{ij}})= \frac{c_{ij} + \gamma_j}{n_i + K_{\backslash i}}$. 
Using a reparametrization of the prior distribution 
 with $\gamma_{\backslash_i}^{\top}= (K_{\backslash_i} (\pi^{\star}_i))^{\top}$ so that the prior expectation can be written as $\E (\underline \theta_i^{\top}) = (\pi_i^{\star})^{\top} $, 
 the posterior expectation \eqref{eq:diricpostexpt}  of the row $i$ can also be written as:
\begin{eqnarray}
\label{eq:diricpostexptbis}
G_i^{{\star}^{\top}} = \alpha_i p_i^{\top} + (1-\alpha_i) (\pi_i^{\star})^{\top}
\end{eqnarray}
with $\pi_i^{\star} = (\pi_{ij}^{\star})_{j\neq i, j =1, \ldots, N}$, $\pi_{ij}^{\star}= \frac{\gamma_j}{K_{\backslash i}}$ and $\pi_{ii}^{\star}=0$.

Equation \eqref{eq:diricpostexptbis} is a classical linear combination of the data $P$ and the prior expectation $\pi_i^{\star}$, which is the prior probability that $i$ cites any other journals.
It has the same form as \eqref{eq:pagerank} so that the modified multinomial-Dirichet model can be considered as an underlying model for EIFA.
However, the damping factor $\alpha$ is no longer fixed, but depends both on the number of references $n_i$ produced by each journal $i$ and on a parameter $K_{\backslash i}$ depending on $i$ and on the parameters $\gamma_j$. The concentration parameter $K_{\backslash i}$, also known as the flattering constant, can be regarded as the total number of "fictive" citations given by $i$. 
The shrinkage coefficient $(1 - \alpha_i)$ given by \eqref{eq:shrink} varies between 0 and 1. It is large when $n_i$ is small and $K_{\backslash i}$ is large.  When the data is not self-sufficient, \textit{i.e.} $K_{\backslash i}$ is large in comparison to $n_i$, there is a need to borrow information from other journals. The limiting case of $n_i=0$ and $\alpha_i = 0$ corresponds to a dangling node. In such a situation, the corresponding row $G_i^{{\star}^{\top}}$ will be equal to the  prior probabilities $(\pi_i^{\star})^{\top}$. This adjustment is automatic contrary to what happens with PageRank as described in the introduction. Finally, we proceed in the same way as page-rank type algorithms to derive a score, \textit{i.e.} we compute the first eigenvector of the matrix $G^{\star}$  \eqref{eq:diricpostexptbis}.

\subsection{Choosing the prior hyperparameters}
\paragraph{Non-informative priors} Various options exist  at this stage to pursue the analysis. The first one corresponds to the situation where the Dirichlet distribution \eqref{eq:diricpost} is completely specified: the parameters are assumed to be known. This corresponds to 
a non-informative  prior distribution  such as
\begin{itemize}
\item Bayes-Laplace: $K = N$ and $\gamma = \mathbf{1}_N$,
\item Jeffreys: $K = N / 2$ and  $\gamma = 1/2 \times \mathbf{1}_N$,
\item Perks: $K = 1$ and $\gamma = 1 / N \times \mathbf{1}_N$.
\end{itemize}
Opinions are divided on the merits of these different priors \citep{Berger,doi:10.1080/00031305.2016.1222309}, especially in the presence of a large number of zero cells, which is precisely the main objective of a PageRank type of smoothing. In all cases, it gives an equiprobability solution to the teleportation $(\pi_i= (1/N, \ldots, 1/N))$ used by PR at the origin. But it was not considered acceptable in bibliometry. 

\paragraph{Empirical Bayes priors}
Another option consists of using  a standard empirical Bayesian framework, also known as the ML-II approach \citep{Berger},  where the hyperparameters $\gamma$ are replaced by the maximum likelihood estimates of the marginal distribution $\law(\underline C |\gamma)$ obtained after integrating out  the parameters of the multinomial distribution $\theta$. The density of this marginal distribution is the product of a compound Dirichlet multinomial (or Polya distribution):

\begin{eqnarray*}
\label{eq:compdiri}
\law(\underline C |\gamma) &=& \Pi_{i=1}^{N} \law(\underline C_i^{\top} |    \gamma_{\backslash i}) \\
\law(\underline C_i^{\top} |    \gamma_{\backslash i})  &=&  \int p(\underline C_i^{\top} |\underline \theta_i^{\top}) p(\underline \theta_i^{\top}|  \gamma_{\backslash i})d\underline \theta_i^{\top} 
 %\int p(\underline C  | \underline \theta) p(\underline \theta | \gamma)  d\underline \theta \\
%\law(\underline C|\gamma) &=&   \\
%&=&
\end{eqnarray*}
where  $p(\underline C_i^{\top} |\underline \theta_i^{\top})$ is the multinomial component:
$$p(\underline C_i^{\top} |\underline \theta_i^{\top}) = \frac{n_i!}{\Pi_{j \neq i} c_{ij}!} \Pi_{j\neq i} \theta_{ij}^{c_{ij}}$$ and    $p(\underline \theta_i^{\top}|  \gamma_{\backslash i}),$ the Dirichlet component:  
 $$p(\underline \theta_i^{\top}|  \gamma_{\backslash i}) =  \frac{\Gamma \left( \sum_{j \neq i} \gamma_j \right)}{\Pi_{j \neq i} \Gamma(\gamma_j)} \Pi_{j \neq i} \theta_{ij}^{\gamma_j-1}.$$
Then, $\Pi_{j \neq i} \theta_{ij}^{c_{ij}+\gamma_j -1}$ is the kernel of a Dirichlet distribution so that $\int \Pi_{j \neq i} \theta_{ij}^{c_{ij}+\gamma_j -1} d \underline \theta_i =  \frac{\Pi_{j \neq i} \Gamma(c_{ij} + \gamma_j)}{\Gamma \left( \sum_{j \neq i} (c_{ij} + \gamma_j)\right)}$
which leads to
\begin{eqnarray}
\label{eq:compdiri}
\law_i(\underline C_i^{\top} |\gamma_{\backslash i}) = \frac{n_i! \Gamma \left( \sum_{j\neq i} \gamma_j \right)} {\Pi_{j \neq i} c_{ij}! \Gamma \left(  \sum_{j \neq i} (c_{ij} + \gamma_j) \right) } \Pi_{j \neq i}\frac{\Gamma (c_{ij} + \gamma_j) }{\Gamma(\gamma_j) } .
\end{eqnarray}

Note that a zero, whether sampling or structural, is not important in the multinomial part, since it multiplies the likelihood by 1 even if we do not exclude the index $i$. However, for the Dirichlet part, it has an impact since we explicitly remove a parameter in each $\theta_i$ so that $\underline \theta_i$ has a size of $(N-1)$ instead of $N$. The impact of such a strategy will be illustrated in Section \ref{sec:appli}.

\subsection{Estimating the hyperparameters in an empirical Bayes framework}

The log-likelihood can be written as:
\begin{eqnarray}
\label{eq:loglik}
L_i(\gamma_{\backslash i}) = \rm{log} \Gamma(K_{\backslash i}) - \rm{log} \Gamma (n_i + K_{\backslash i}) + \sum_{j \neq i} \left[ log \Gamma(c_{ij} + \gamma_j) - \rm{log} \Gamma (\gamma_j) \right], 
\end{eqnarray}
and its gradient can be written as:
\begin{eqnarray}
\label{eq:grad}
\frac{d \law_i(\underline C_i^{\top}  | \gamma_{\backslash i})  }{ d \gamma_j}&=&   \psi(\sum_{j \neq i} \gamma_j ) +   \psi(n_i+\sum_{j \neq i} \gamma_j  )  +  \psi(c_{ij} + \gamma_j) - \psi(\gamma_j), \mbox{ for all } i\neq j\\
 \frac{d \law_i(\underline C_i^{\top}  | \gamma_{\backslash i})  }{ d \gamma_i} &=& 0.
\end{eqnarray}
We then sum to get:
\begin{eqnarray}
\label{eq:gradind}
[\nabla L(\gamma)]_{j} =\frac{d \law (\underline C | \gamma)}{d\gamma_j} =    \sum_{i \neq j} \frac{d \law (\underline C_i | \gamma_{\backslash i})}{d\gamma_j} =  \sum_{i \neq j}  \psi(K_{\backslash i}) - (N-1) \psi(\gamma_j) + \sum_{i \neq j} \left( \psi(c_{ij}+ \gamma_j) - \psi(n_i + K_{\backslash i}) \right), 
\end{eqnarray}
where $\psi(x) = d \rm{log} \Gamma(x)/dx$, is the digamma function.

Different algorithms can be used to maximize the log-likelihood \eqref{eq:loglik} such as Minorization-Maximiza\-tion algorithms. Indeed, using results from \citet{Minka2012}, we get a lower bound for the likelihood that can be iteratively maximized. It leads to a fixed point iteration algorithm 
%The gradient vector of the loglikelihood $\nabla L(\gamma) = \partial L(\gamma) / \partial %\gamma$ %results in the fixed point iteration algorithm 
defined for $j=1, \ldots, N$, by iterating:                       
% pour trouver f(x)=0, on ecrit x=g(x) et on itère ça.
\begin{eqnarray}
\label{eq:fp}
 \gamma_{j}^{\ell+1}  = \gamma_{j}^\ell \frac{\sum_{i \neq j} \psi (c_{ij}  + \gamma_{j}^{\ell}) - (N-1) \psi(\gamma_j^{\ell})}{\sum_{i \neq j} \psi (n_i  + K_{\backslash i}^{\ell}) -  \psi( K_{\backslash i}^{\ell})}.
\end{eqnarray}

Other options include a first order algorithm with inversion, second order algorithms such as Levenberg-Marquardt which require computing the Hessian matrix, or an expectation-maximization (EM) variant of this latter algorithm.  Of course, second order algorithms have the great advantage of giving as an output the asymptotic sampling variance-covariance matrix but this comes at the price of supplement burden from the computational point of view.  
In the Appendix, % (\ref{section:append}), 
we detail the other algorithms and give a small comparison study highlighting the merits of the first order fixed point method.  %that we use in the sequel.

\textbf{Remark:} 
Denoting $\pi_j=\frac{\gamma_j}{K}$, a simpler empirical Bayesian strategy could be to fix $\pi$ either as $\pi=(\tilde a_i)_{i=1, \ldots, N}$ or 
 $\pi=(c_{+i}/c_{++})_{i=1, \ldots, N}$ and to replace $K$ with its maximum likelihood estimation.   This plug-in value can be obtained with a single fixed-point equation leading to iterating:
\begin{eqnarray}
\label{eq:K}
 K^{\ell+1}  = K^\ell \frac{\sum_{i=1}^N\sum_{i \neq j}   \pi_j \psi (c_{ij}  + K^{\ell} \pi_{j}) - (N-1) \psi(K^{\ell} \pi_j)}{\sum_{i=1}^N(1-\pi_i)\psi (n_i  + K^{\ell}(1-\pi_i)) -  \psi( K^{\ell}(1-\pi_i))}.
\end{eqnarray}
%An even simpler  solution would be to fix  $\pi$ as before  and setting $K=N$. This would make EBEF and EIFA closer to each other and EBEF could be easier to compute.  

\section{Ranking statistical journals} \label{sec:appli}

The application concerns the matrix $C$ of cross-references between 47 statistical journals, studied by \citet{CIS-448159}. It concerns citations published in 2010 related to articles published from 2001 to 2010.  A subset of the matrix is given in Table \ref{tab:extra} and the complete list with abbreviations is given in Table \ref{liste47} of Appendix \ref{app:rank} .
\begin{table}[ht]
\small
\centering
\begin{tabular}{rlrrrrr}
  \hline
 &  & AmS & AISM & AoS & ANZS & Bern \\ 
  \hline
1&AmS &  43 &   0 &   9 &   0 &   1 \\ 
  2&AISM &   1 &  18 &  24 &   5 &   7 \\ 
  3&AoS &   2 &   3 & 291 &   2 &  27 \\ 
  4&ANZS &   0 &   3 &   4 &   5 &   0 \\ 
  5&Bern &   0 &   5 &  53 &   0 &  22 \\ 
   \hline
\end{tabular}
\caption{Extract of the 47 $\times$ 47 cross-citation matrix between statistics journals.}
\label{tab:extra}
\end{table}
We apply our method described in Section \ref{section:ebay} called EBEF for Empirical Bayes Eigen Factor.
The maximum likelihood estimate and the associated variance of the parameter $K$ of concentration is established at $K$ = 58.10  +/- 2.82 with a significant variation between the  $\gamma_j$ values ranging from 6.61 +/- 0.54 for JASA to 0.06 +/- 0.03 for STATAJ. Note that ignoring
self-citations by considering these data as sampling zeros (using a standard Dirichlet-multinomial  scheme without specific modification for the Dirichlet parameters as in equation \eqref{eq:diric}) leads to a substantially  different estimate of $K$ ($K$ = 49.00). 

Looking at the parameters $\alpha_i$ shows that journals such as CSDA or STMED have values  close to 0.95 whereas  STATAJ has a value of 0.39  (the mean is 0.77). Teleportation is  decreasing with the number of references emitted by a journal. It makes sense since it is less relevant to use teleportation in the case of many outgoing links. 
The scores obtained by EBEF are given in Table \ref{fig:ranktotal} column 4 and associated ranks are given in Table \ref{fig:ranktotalR} in Appendix \ref{app:rank}. %It shows that EBEF favors journals which publish a lot of papers.

\begin{table}[!h]
\centering
\scriptsize
\begin{tabular}{rlrrrr}
  \hline
 &Journal & PSJR & EBPR & EIFA & EBEF \\ 
  \hline
1 & JASA & 119.51 & 132.49 & 126.84 & 127.28 \\ 
  2 & AOS & 105.35 & 116.95 & 97.62 & 97.17 \\ 
  3 & JRSS-B & 71.12 & 79.67 & 78.24 & 79.91 \\ 
  4 & BKA & 62.43 & 68.48 & 71.92 & 72.97 \\ 
  5 & BCS & 63.11 & 66.57 & 64.23 & 63.81 \\ 
  6 & STMED & 67.01 & 62.17 & 53.63 & 51.27 \\ 
  7 & JSPI & 45.88 & 40.46 & 43.76 & 42.39 \\ 
  8 & CSDA & 45.71 & 42.24 & 38.16 & 38.44 \\ 
  9 & STSIN & 28.09 & 29.63 & 33.69 & 34.45 \\ 
  10 & JMA & 29.96 & 29.26 & 30.74 & 30.27 \\ 
  11 & BIOST & 23.65 & 25.01 & 26.77 & 26.65 \\ 
  12 & JCGS & 21.69 & 23.33 & 24.28 & 25.10 \\ 
  13 & SPL & 31.90 & 23.89 & 24.78 & 23.75 \\ 
  14 & SJS & 18.34 & 19.66 & 22.73 & 23.47 \\ 
  15 & STSCI & 18.70 & 20.93 & 22.81 & 23.26 \\ 
  16 & BERN & 14.77 & 14.65 & 16.70 & 16.31 \\ 
  17 & CJS & 11.09 & 11.80 & 13.16 & 13.79 \\ 
  18 & STCMP & 11.19 & 11.89 & 12.87 & 13.42 \\ 
  19 & BIOJ & 11.98 & 11.04 & 12.14 & 12.14 \\ 
  20 & TECH & 11.76 & 11.60 & 11.65 & 11.80 \\ 
  21 & CSTM & 17.59 & 10.74 & 13.50 & 11.61 \\ 
  22 & JRSS-C & 9.30 & 9.55 & 10.51 & 11.08 \\ 
  23 & TEST & 7.94 & 8.51 & 9.87 & 10.27 \\ 
  24 & JRSS-A & 10.93 & 10.44 & 9.66 & 9.81 \\ 
  25 & AISM & 9.73 & 8.66 & 10.28 & 9.79 \\ 
  26 & AMS & 10.28 & 9.82 & 9.17 & 9.62 \\ 
  27 & JNS & 7.82 & 7.25 & 8.59 & 8.54 \\ 
  28 & LTA & 7.67 & 7.61 & 8.60 & 8.52 \\ 
  29 & JSCS & 8.03 & 6.43 & 7.18 & 7.31 \\ 
  30 & ENVR & 7.90 & 6.88 & 6.91 & 7.26 \\ 
  31 & SMMR & 6.15 & 5.86 & 6.67 & 6.54 \\ 
  32 & MTKA & 6.35 & 5.11 & 6.21 & 5.77 \\ 
  33 & CSSC & 7.32 & 4.77 & 6.08 & 5.63 \\ 
  34 & JSS & 6.52 & 6.13 & 5.17 & 5.43 \\ 
  35 & JTSA & 6.81 & 5.28 & 5.88 & 5.43 \\ 
  36 & ANZS & 4.84 & 4.38 & 5.02 & 5.28 \\ 
  37 & JBS & 7.22 & 5.45 & 5.35 & 5.12 \\ 
  38 & STATS & 4.92 & 4.13 & 5.21 & 5.04 \\ 
  39 & ISR & 5.35 & 4.91 & 4.86 & 5.03 \\ 
  40 & JAS & 6.58 & 4.09 & 4.83 & 4.72 \\ 
  41 & CMPST & 4.18 & 3.71 & 4.12 & 4.49 \\ 
  42 & JABES & 4.46 & 4.01 & 4.26 & 4.44 \\ 
  43 & STMOD & 3.56 & 3.60 & 3.95 & 4.28 \\ 
  44 & STNEE & 3.41 & 3.16 & 3.45 & 3.83 \\ 
  45 & EES & 4.41 & 3.75 & 3.41 & 3.55 \\ 
  46 & STPAP & 3.45 & 2.09 & 2.70 & 2.45 \\ 
  47 & STATAJ & 4.27 & 2.18 & 2.07 & 1.73 \\ 
   \hline
\end{tabular}
\caption{\label{fig:ranktotal}\footnotesize{Total influence scores. PSJR: Prestige Scimago Journal Rank (self-citations restricted to 33\% of references); EBPR: Empirical Bayes PageRank (self-citations included);
EIFA*: Eigenfactor; EBEF*: Empirical Bayes Eigen Factor
* without self-citations. Journals are ordered according to EBEF scores. } } \label{fig:ranktotal}
\end{table}

%PageRank has a size bias: larger journals or fields
%receive more citations overall,so will have larger
%PageRank.
%This is a problem if we want to measure prestige,
%rather than popularity.
%The Scroogefactor, defined as PageRank per
%outgoing citation, controls for this: journals
%are, in effect, penalised for being generous with
%citations and rewarded for being miserly.
%When the citation matrix is quasi-symmetric,
%the Scroogefactor is exactly equal to the
%Bradley–Terry scores

We compare EBEF to EIFA  (column 3) since it can be regarded as its Bayesian counterpart. Note that here EIFA is applied on a dataset where 10 years are considered whereas the official time window is 5 years.
We also add two scores that include  self-citations namely EBPR (Empirical Bayes Page Rank) and the Prestige Scimago Journal Rank (PSJR). 
EBPR  consists of using the Dirichlet-multinomial scheme but without excluding the diagonal. The implementation is straightforward from EBEF. 
The PSJR score is produced by Scimago Lab and released by Scopus, the citation database of Elsevier. 
It is defined as 
\begin{eqnarray}
\label{eq:PSJR}
G_2 = \alpha_2 P+ (1-\alpha_2 - \beta)\mathbf{1} \pi^{\top} + \beta \frac{\mathbf{1} \mathbf{1}^{\top} }{N}, 
\end{eqnarray}
with $\pi=(\tilde a_i)_{1, \ldots, N}$ where $\tilde a_i = a_i/a_{+}$, $\alpha_2= 0.90$ and $\beta=10^{-4}$.  (Recall that $a_i$ is the number of articles published 
by $i$ in the considered time window.) Then the first eigenvector is computed: $G_2^{\top} r_2=r_2$. PSJR uses a teleportation which depends on the number of papers published and adds a small term of uniform teleportation. In addition, self-citations are restricted to 33\% of references emitted by each journal. 
We apply \eqref{eq:PSJR} on the  same cross-references matrix between the 47 journals for fair comparisons. 

However, PSJR uses the information contained in the Scopus data base of
Elsevier on a 3-year window while EIFA is based on journals indexed by
Clarivate Analytics (ex Thomson-Reuters) on a 5-year window.
This makes the comparison of both scores EIFA and PSJR, released by two concurrents, more difficult. 

To get rid of the strong effect of the number of papers published, we favor the scores normalized with the number of published papers, exactly as the article influence (AI) discussed in the introduction. It gives sensitive different results as illustrated in Table \ref{fig:rankarticle} for the scores and in Table \ref{fig:rankarticleR} (see Appendix \ref{app:rank}) for the ranks.
\begin{table}[!ht]
\centering
\scriptsize
\begin{tabular}{rlrrrr}
  \hline
 & Journal &PSJR &  EBPR & EIFA (AI) & EBEF \\ 
  \hline
1 & JRSS-B & 5.31 & 5.95 & 5.84 & 5.97 \\ 
  2 & STSCI & 3.46 & 3.87 & 4.22 & 4.30 \\ 
  3 & AOS & 4.12 & 4.58 & 3.82 & 3.80 \\ 
  4 & JASA & 3.44 & 3.81 & 3.65 & 3.66 \\ 
  5 & BKA & 2.66 & 2.92 & 3.07 & 3.12 \\ 
  6 & BIOST & 1.71 & 1.81 & 1.93 & 1.92 \\ 
  7 & SJS & 1.40 & 1.50 & 1.74 & 1.80 \\ 
  8 & JCGS & 1.52 & 1.63 & 1.70 & 1.76 \\ 
  9 & BCS & 1.70 & 1.80 & 1.73 & 1.72 \\ 
  10 & STSIN & 1.28 & 1.35 & 1.53 & 1.56 \\ 
  11 & TEST & 1.15 & 1.24 & 1.44 & 1.49 \\ 
  12 & CJS & 1.15 & 1.22 & 1.36 & 1.43 \\ 
  13 & STCMP & 1.11 & 1.18 & 1.28 & 1.33 \\ 
  14 & BERN & 1.05 & 1.04 & 1.19 & 1.16 \\ 
  15 & TECH & 1.05 & 1.04 & 1.04 & 1.06 \\ 
  16 & LTA & 0.94 & 0.93 & 1.05 & 1.04 \\ 
  17 & JRSS-C & 0.79 & 0.82 & 0.90 & 0.95 \\ 
  18 & JRSS-A & 1.03 & 0.98 & 0.91 & 0.93 \\ 
  19 & JMA & 0.89 & 0.87 & 0.91 & 0.90 \\ 
  20 & STMOD & 0.64 & 0.65 & 0.71 & 0.78 \\ 
  21 & SMMR & 0.70 & 0.66 & 0.76 & 0.74 \\ 
  22 & AMS & 0.76 & 0.72 & 0.68 & 0.71 \\ 
  23 & ISR & 0.74 & 0.68 & 0.68 & 0.70 \\ 
  24 & AISM & 0.65 & 0.58 & 0.69 & 0.65 \\ 
  25 & JNS & 0.59 & 0.54 & 0.64 & 0.64 \\ 
  26 & STMED & 0.82 & 0.76 & 0.66 & 0.63 \\ 
  27 & BIOJ & 0.58 & 0.54 & 0.59 & 0.59 \\ 
  28 & CSDA & 0.66 & 0.61 & 0.55 & 0.55 \\ 
  29 & JSS & 0.66 & 0.62 & 0.52 & 0.55 \\ 
  30 & JSPI & 0.58 & 0.51 & 0.56 & 0.54 \\ 
  31 & JABES & 0.50 & 0.45 & 0.48 & 0.50 \\ 
  32 & ANZS & 0.46 & 0.41 & 0.47 & 0.50 \\ 
  33 & STNEE & 0.41 & 0.38 & 0.41 & 0.46 \\ 
  34 & STATS & 0.44 & 0.37 & 0.46 & 0.45 \\ 
  35 & EES & 0.55 & 0.47 & 0.42 & 0.44 \\ 
  36 & MTKA & 0.48 & 0.38 & 0.47 & 0.44 \\ 
  37 & JTSA & 0.54 & 0.42 & 0.47 & 0.43 \\ 
  38 & ENVR & 0.46 & 0.40 & 0.41 & 0.43 \\ 
  39 & CMPST & 0.37 & 0.33 & 0.37 & 0.40 \\ 
  40 & JSCS & 0.36 & 0.29 & 0.32 & 0.33 \\ 
  41 & SPL & 0.39 & 0.30 & 0.31 & 0.29 \\ 
  42 & JBS & 0.40 & 0.31 & 0.30 & 0.29 \\ 
  43 & CSSC & 0.28 & 0.18 & 0.23 & 0.21 \\ 
  44 & CSTM & 0.30 & 0.18 & 0.23 & 0.20 \\ 
  45 & STATAJ & 0.46 & 0.24 & 0.22 & 0.19 \\ 
  46 & STPAP & 0.26 & 0.16 & 0.20 & 0.18 \\ 
  47 & JAS & 0.25 & 0.15 & 0.18 & 0.18 \\ 
   \hline
\end{tabular}
\caption{\footnotesize{Article influence scores. PSJR: Prestige Scimago Journal Rank (self-citations restricted to 33\% of references); EBPR: Empirical Bayes PageRank (self-citations included);
EIFA*: Eigenfactor; EBEF*: Empirical Bayes Eigen Factor
* without self-citations. Journals are ordered according to EBEF scores.} } \label{fig:rankarticle}
\end{table}
Indeed, with the EBEF, the rank of CSDA decreases from 8 to 28. On the contrary TEST increases from rank 23 to 11, etc. 
The top quintet,  JRSS-B, STSCI, AOS, JASA, BKA is stable across the article influence scores. 
%Another solution consists of using scores normalized with the number of emited citations. % This is illustrated in the Appendix \ref{app:rank} in Table \ref{fig:rankcitat}.

In addition, we compute in Table \ref{fig:corrscores} both Spearman rank correlation and Kendall tau correlation to assess the similarities between rankings given by the different scores. EBEF is highly correlated with EIFA which was expected thus giving credit to EIFA for its efficiency. The same applies to PSJR with respect to EBPR.  
Although highly correlated (kendal-tau of 0.90), EBPR and EBEF do not rate and rank journals in the same way, making clear the importance of self-citations in defining journal influence rating systems. 

\begin{table}[!ht]
\centering
\small
\begin{tabular}{rllll}
  \hline
 &   PSJR & EBPR & EIFA & EBEF \\ 
  \hline
PSJR & 1 & 0.994 & 0.978 & 0.9753 \\
EBPR & 0.964 & 1 & 0.987 & 0.987 \\
EIFA & 0.893 & 0.92 1& 1 & 0.996 \\
EBEF & 0.886 & 0.918 & 0.965 & 1\\
   \hline
\end{tabular}
\caption{Correlations among article based journal scores. Below diagonal: Kendall tau. Above diagonal: Spearman rank correlation. } \label{fig:corrscores}
\end{table}

%The multiplication in EIFA by the initial matrix $P$ without self-citation seems to remove the %part of the effect of the smoothing with additional diagonal terms [A bit obverstatted?]. 

\section{Discussion}

The suggested method EBEF is an extension of the PR-type algorithm used for the establishment of the Eigenfactor according to a well-established probabilistic model (Dirichlet-multinomial model) which 
allows the constraint of  exclusion of self-citations to be treated in a rigorous manner. The smoothing of the adjacent matrix corresponding to the citing to cited network is obtained as in PR by a convex combination of the corresponding vector of the observed transition probabilities and of a teleportation vector according to respective probabilities which vary from one journal to another as a function of the total number of references $n_i$ and of a concentration coefficient $K_{\backslash i}$. In addition to its conceptual and computational simplicity, the Bayesian PR developed here has the merit of taking into account and distinguishing the zeros of structure from those of sampling.
The way we deal with structural zeroes can be extended to non-diagonal terms to take into account other constraints such as restriction or exclusion of links between subgroups of journals belonging to a specific field.

The maximum likelihood estimate of the parameters, derived with a Majorize-Minimize algorithm, could also be obtained as a by-product of hierarchical Bayes strategy with an additional step of specifying  non-informative prior distributions on the parameters $\gamma$s and computing the  posterior modes of the corresponding marginal distributions. 

Finally, one potential concern with the empirical Bayes approach is that it uses the data  twice, both to estimate the parameters $\gamma$s and to compute the final scores and ranking. To tackle this issue, one can resort to a Monte Carlo half sampling procedure as follows. 
A training matrix $\tilde C = (\tilde c_{ij})$  is generated with $\tilde c_{ij} \sim  \mbox{Binomial}(c_{ij}, (1-\delta))$, with $\delta=0.5$. 
Parameters $\gamma$s are estimated with the training matrix and these estimates are used to calculate PSJR,
EBPR, EIFA and EBEF scores on the complementary matrix $C -\tilde C$. This process is repeated
$m=200$ times and final scores are obtained as means of elementary scores over
the $m$ replications. More details about this procedure,  extensions and results are given in in Appendix \ref{app:half}.

Dealing with self-citations is a critical topic since including or excluding it may appear too radical. The \textit{ad hoc} solution used by PSJR which consists of bounding the self-citations to 33\% could be a solution. We suggest an alternative solution allowing  underweighting self-citations in a data-driven way.
It is as follows. 
Let us consider the simple score defined for journal $i$ as the ratio of the total number of
citations received ($c_{i+}= \sum_{j=1}^N c_{ij}$)
 by $i$ from other journals including itself divided by the total
number of references made by this to other journals: $S_i=\frac{c_{+i}}{c_{i+}}$. 
This ratio provides a natural starting approximation to the iterative algorithm for computing the
"influence weight" of journal $i$ introduced by \citet{pinskin76}. The numerator of $S_i$ can 
be decomposed into self ($c_{ii}$) and external ($R_{\backslash i}$) citations received and the denominator
likewise into self ($c_{ii}$) and external ($M_{\backslash i}$) references made so that:
\begin{equation*}
S_i=\frac{c_{+i}}{c_{i+}} = \frac{c_{ii}+R_{\backslash i}}{c_{ii}+M_{\backslash i} }.
\end{equation*}
Let $\kappa \in [0,1]$ be a tuning parameter devoted to attenuate the effect of self-citations. Define
\begin{equation*}
S_i(\kappa)=\frac{\kappa c_{ii}+R_{\backslash i}}{\kappa c_{ii}+M_{\backslash i}}.
\end{equation*}
It can easily be shown that 
\begin{itemize}
\item if $S_i (0)< 1$, $S_i(\kappa)$ is an increasing function of $\kappa$ which remains upper bounded by 1.
\item  if $S_i (0)=1$, $S_i(\kappa)=1$ for any $\kappa$.
\item if $S_i (0)>1$, $S_i(\kappa)$ is a decreasing function of $\kappa$ which remains lower bounded by 1.
\end{itemize}
This scores implies that powerful journals have no interest in favoring self-citations contrary to journals of lower status.
However, the impact of self-citations remains bounded.
Based on this comment, we would suggest selecting $\kappa$ for journal $i$, with 
\begin{equation*}
\kappa_i=\min \left(\frac{\min(R_i,M_i)}{c_{ii}},1\right),
\end{equation*}
which  penalizes lower status journals trying to take advantage of self-citations. 

Applying this rule  to the 47 statistical journals at hand in this paper leads to the choice of $\kappa_i=1$ for all the journals except for STATATJ
which receives the weight $\kappa=0.442$ (self-citation rate= 67\%) and JSS which receives the weight $\kappa=0.887$ (self-citation rate= 32\%). These journals have by their very nature a lot of self-citations since they are  related to softwares. In particular, STATAJ is exclusively the journal of the software STATA. There are two other journals having a self-citation rate greater than the PSJR threshold of 33\%:  Annals of Statistics (36\%) and Statistics in Medecine (37\%) but still with $\kappa$ values of 1.
This choice of weights seems quite sensible. Actually, most of the statistical journals do not use much self-citations (the mean of self-citation rate is 20\%). 

%favor nor encourage self-citations as a good research publishing practice except when necessary due to highly specialized fields covered by very few journals.

\section{Appendix}

\subsection{Algorithms comparison} \label{section:append}

In this section, we describe and compare other algorithms that can be used to maximize the marginal likelihood \eqref{eq:loglik}. 
From \eqref{eq:gradind}, we can write
\begin{eqnarray}
\label{eq:fornewt}
\psi(\gamma_j) =   \frac{1}{(N-1)} \left[ \sum_{i \neq j}  \psi(K_{\backslash i})  + \sum_{i \neq j} \left( \psi(c_{ij}+ \gamma_j) - \psi(n_i + K_{\backslash i}) \right) \right].
\end{eqnarray}
Formula \eqref{eq:fornewt} serves as a basis for an inversion method (INV) i.e. $\psi(\gamma_j)=a$ that can be solved via Newton Raphson by iterating: 
\begin{eqnarray} \label{eq:INV}
\gamma_j^{\ell+1} = \gamma_j^{\ell} - \frac{\psi(\gamma_j^{\ell})  - a^{\ell} }{\psi^{'}(\gamma_j^{\ell})},
\end{eqnarray}
with $a^{\ell}= (N-1)^{-1}  \left[\sum_{i \neq j} \left(  \psi(K_{\backslash i}^{\ell}) - \psi(n_i+ K_{\backslash i}^{\ell})  \right)  + \sum_{i \neq j} \psi(c_{ij} + \gamma_j^{\ell}) \right]$. \\

Second order algorithms can also be considered. 
The second derivatives can be written as: 
\begin{eqnarray}
\frac{d^2 \law (\underline C | \gamma)}{d\gamma_j^2} =  \sum_{j \neq i}  \frac{d^2 \law (\underline C_i | \gamma_{\backslash i})}{d\gamma_j^2}= 
\sum_{i \neq j}  \psi^{'}(K_{\backslash i}) - (N-1) \psi^{'}(\gamma_j) + \sum_{i \neq j} \left( \psi^{'}(c_{ij}+ \gamma_j) - \psi^{'}(n_i + K_{\backslash i}) \right), 
\end{eqnarray}
with $\psi^{'}$ the trigamma function
and 
\begin{eqnarray}
\frac{d^2 \law (\underline C | \gamma)}{d\gamma_j d \gamma_k} =  \sum_{j \neq i \neq k}  \frac{d^2 \law (\underline C_i | \gamma_{\backslash i})}{d\gamma_j d \gamma_k}= 
\sum_{i \neq j \neq k}  \psi^{'}(K_{\backslash i})  - \psi^{'}(n_i + K_{\backslash i}). 
\end{eqnarray}
Denoting the Hessian matrix as $H(\gamma)= \frac{d^2L(\gamma)}{d\gamma d \gamma^{\top}}$, a Levenberg-Marquardt algorithm consists of iterating:
\begin{eqnarray} \label{eq:LM}
\left[H(\gamma^{\ell}) + \lambda^{\ell} \rm{diag} \left( H(\gamma^{\ell}) \right) \right] \left( \gamma^{\ell+1}  - \gamma^{\ell} \right) = \nabla L(\gamma^{\ell}),
\end{eqnarray}
where $\lambda^{\ell}$ is a damping factor adjusted at each iteration with decreasing values if $L(\gamma)$ increases and of increasing values if $L(\gamma)$ decreases. When $\lambda^{\ell}=0$, the algorithm boils down to Newton-Raphson. The adaptive sequence $(\lambda_{1}, \ldots, \lambda_{L})$, with $L$ the number of iterations, can be chosen as suggested by \citet{8f6705d73b514dd58e9323a7a4ff0eaf} and \citet{doi:10.1080/00949655.2016.1238086} with:
\begin{eqnarray}
\lambda^{\ell+ 1} = \lambda^{\ell} \max \left(1/3, 1-(2 \rho^{\ell} -1)^3 \right) \mbox{ if } \rho^{\ell+1}>0, \\
\lambda^{\ell+ 1} = 2\lambda^{\ell} \mbox{ otherwise}
\end{eqnarray} 
with, $\lambda^{0}=0$ and $\rho^{\ell+1} = \frac{L(\gamma^{\ell+1}) - L(\gamma^{\ell}) )}{ 1/2 (  \gamma^{\ell+1}- \gamma^{\ell} )^{\top} H(\gamma^{\ell}) (  (  \gamma^{\ell+1}- \gamma^{\ell} )) }$. The stopping rule can be defined as $\frac{|| \gamma^{\ell+1}- \gamma^{\ell}  ||}{ ||\gamma^{\ell}   || + \varepsilon_1} < \varepsilon_2$.

It is also possible to derive an EM algorithm to maximize the likelihood with $\theta$ being regarded as the missing latent variables. However, the EM is not relevant since a first order algorithm to achieve the M step leads to equation \eqref{eq:grad}. 

 %One then has to maximize the following function:
%C denotes a conditional expectation with respect to the distribution of? Knowing the data C and the current values ??of the %parameters (t)? . The derivative of (14) has a large
%similarity
%[QUESTION: A t'on un maximum unique probleme convex? quelles sont les propriétés de %Leven Mark et peut-on utiliser des algos de quasi-Newton?  ecrit-on EM?]
%In order to check the sensitivity of the algorithms to initial conditions, different starting points were considered: $a) \gamma_j^{0}=(1/N)$ (Perks), $b) \gamma_j^{0}=1$ (Bayes-Laplace) and $c) \gamma_j^{0} = N c_{+j}/c_{++}$. 
\begin{table}[ht]
\centering
\small
\begin{tabular}{l llllllllllll}
  \hline
 & FP& & & INV &&& LM &&& LMem&& \\ 
 & a &b& c & a & b &c & a &b &c & a &b & c \\
Nb iterations & 86 &75& 54 & 88 &99 &66 & 15& 11& 8 & 61 &45 &44 \\
Times (s) &  6& 4 &3 & 9 &11 &8 & 37 &26 &25 \\
   \hline
& FP& & & INV &&& LM &&& LMem&& \\ 
 & a& b& c & a& b& c & a& b& c & a& b&  c \\
Nb iterations & 107 &96& 75 & 111 &127& 84 & 16& 12 &9 & 87& 70 &69 \\
Times (s) &  7 &6 &5 & 12& 13 &9 & 17 &12 &9 & 43 &41 &40 \\
\end{tabular}
\caption{Comparison of algorithms used to compute maximum-likelihood estimations of the Compound
Dirichlet Multinomial parameters. FP: Fixed Point Iteration; INV: Inversion Method; LM: Levenberg-Marquardt; LMem: LM for EM. Top table with stopping threshold equals to $\varepsilon_2 = 10^{-5}$ and bottom table $\varepsilon_2 = 10^{-6}$.} \label{fig:algo}
\end{table}

As shown in Table \ref{fig:algo}, there is a striking contrast between the performance of the algorithms
in terms of number of iterations \textit{versus} computing time to convergence. The Levenberg-Marquardt (LM) algorithm \eqref{eq:LM} is by far the algorithm needing the lowest number of iterations
whatever the stopping rule and the starting point. 
Fixed point iteration (FP)  \eqref{eq:fp} and inversion (INV) \eqref{eq:INV} require 7 to 8 times more runs for
convergence. But, as far as computing time is concerned, the most efficient algorithm is FP
(5 to 7 seconds for $\varepsilon_2 =10^{-6}$) while INV and LM require twice as much time but within
reasonable figures. Moreover, LM produces an estimate of the asymptotic sampling
variance-covariance matrix whereas FP and INV do not.
In conclusion, one may suggest using FP to get a quick estimation of parameters and then
checking it with LM. Incidentally, the EM-based LM algorithm does not display any
advantage as compared to the standard LM. In all cases, a start with the empirical value  $\gamma_j^{0} = N c_{+j}/c_{++}$
provides the most effective performance both in time and number of
iterations, but using the other ones is not worthless to check insensitivity of the solutions to
initial conditions.
%Final maximum likelihood estimations are shown in Table \ref{gamma}. 

There is a strong correlation (0.967) between the
ML estimates of the  parameters and the numbers or proportions of cites received by
the different journals with the highest values for JASA, AOS, JRSS-B, BKA, BCS and the lowest
for STATJ, EES, JBS, STPAP, JABES confirming that it 
is a good starting value $\gamma_j^{0}$. However, the concentration parameter $K$ remains a key issue in the estimation
process. For the starting values we took $K = N = 47$ whereas the estimate turns out to be 58 +-2.82. 

\subsection{Ranking statistical journals} \label{app:rank}
Table \ref{liste47} displays the list of the 47 statistical journals.
\begin{table}[ht]
\scriptsize
\centering
\begin{tabular}{rll}
  \hline
 & Journal.Name & Abbreviation \\ 
  \hline
1 & American Statistician & AmS \\ 
  2 & Annals of the Institute of Statistical Mathematics & AISM \\ 
  3 & Annals of Statistics & AoS \\ 
  4 & Australian and Neww Zealand Journal of Statistics & ANZS \\ 
  5 & Bernoulli & Bern \\ 
  6 & Biometrical Journal & BioJ \\ 
  7 & Biometrics & Bcs \\ 
  8 & Biometrika & Bka \\ 
  9 & Biostatistics & Biost \\ 
  10 & Canadian Journal of Statistics & CJS \\ 
  11 & Communication in Statistics-Simulation and Computation & CSSC \\ 
  12 & Communication in Statistics-Theory and Methods & CSTM \\ 
  13 & Computational Statistics & CmpSt \\ 
  14 & Computational Statistics and Data Analysis & CSDA \\ 
  15 & Environmental and Ecological Statistics & EES \\ 
  16 & Environmetrics & Envr \\ 
  17 & International Statistical Review & ISR \\ 
  18 & Journal of Agricultural Biological and Environmental Statistics & JABES \\ 
  19 & Journal of the American Statistical Association & JASA \\ 
  20 & Journal of Applied Statistics & JAS \\ 
  21 & Journal of Biopharmaceutical Statistics & JBS \\ 
  22 & Journal of Computational and Graphical Statistics & JCGS \\ 
  23 & Journal of Multivariate Analysis & JMA \\ 
  24 & Journal of Nonparametric Statistics & JNS \\ 
  25 & Journal of the Royal Statistical Society, Series A & JRSS-A \\ 
  26 & Journal of the Royal Statistical Society, Series B & JRSS-B \\ 
  27 & Journal of the Royal Statistical Society, Series C & JRSS-C \\ 
  28 & Journal of Statistical Computation and Simulation & JSCS \\ 
  29 & Journal of Statistical Planning and Inference & JSPI \\ 
  30 & Journal of Statistical Software & JSS \\ 
  31 & Journal of Time Series Analysis & JTSA \\ 
  32 & Life Data Analysis & LDA \\ 
  33 & Metrika & Mtka \\ 
  34 & Scandinavian Journal of Statistics & SJS \\ 
  35 & Stata Journal & StataJ \\ 
  36 & Statistics and Computing & StCmp \\ 
  37 & Statistics & Stats \\ 
  38 & Statistics in Medicine & StMed \\ 
  39 & Statistical Methods in Medical Research & SMMR \\ 
  40 & Statistical Modelling & StMod \\ 
  41 & Statistica Neerlandica & StNee \\ 
  42 & Statistical Papers & StPap \\ 
  43 & Statistics and Probability Letters & SPL \\ 
  44 & Statistical Science & StSci \\ 
  45 & Statistica Sinica & StSin \\ 
  46 & Technometrics & Tech \\ 
  47 & Test & Test \\ 
   \hline
\end{tabular}
\caption{List and abbreviations of the 47 statistical Journals}\label{liste47}
\end{table}

\begin{table}[ht]
\centering
\scriptsize
\begin{tabular}{rllll}
  \hline
 & PSJR & EBPR & EIFA & EBEF \\ 
  \hline
1 & JASA & JASA & JASA & JASA \\ 
  2 & AOS & AOS & AOS & AOS \\ 
  3 & JRSS-B & JRSS-B & JRSS-B & JRSS-B \\ 
  4 & STMED & BKA & BKA & BKA \\ 
  5 & BCS & BCS & BCS & BCS \\ 
  6 & BKA & STMED & STMED & STMED \\ 
  7 & JSPI & CSDA & JSPI & JSPI \\ 
  8 & CSDA & JSPI & CSDA & CSDA \\ 
  9 & SPL & STSIN & STSIN & STSIN \\ 
  10 & JMA & JMA & JMA & JMA \\ 
  11 & STSIN & BIOST & BIOST & BIOST \\ 
  12 & BIOST & SPL & SPL & JCGS \\ 
  13 & JCGS & JCGS & JCGS & SPL \\ 
  14 & STSCI & STSCI & STSCI & SJS \\ 
  15 & SJS & SJS & SJS & STSCI \\ 
  16 & CSTM & BERN & BERN & BERN \\ 
  17 & BERN & STCMP & CSTM & CJS \\ 
  18 & BIOJ & CJS & CJS & STCMP \\ 
  19 & TECH & TECH & STCMP & BIOJ \\ 
  20 & STCMP & BIOJ & BIOJ & TECH \\ 
  21 & CJS & CSTM & TECH & CSTM \\ 
  22 & JRSS-A & JRSS-A & JRSS-C & JRSS-C \\ 
  23 & AMS & AMS & AISM & TEST \\ 
  24 & AISM & JRSS-C & TEST & JRSS-A \\ 
  25 & JRSS-C & AISM & JRSS-A & AISM \\ 
  26 & JSCS & TEST & AMS & AMS \\ 
  27 & TEST & LTA & LTA & JNS \\ 
  28 & ENVR & JNS & JNS & LTA \\ 
  29 & JNS & ENVR & JSCS & JSCS \\ 
  30 & LTA & JSCS & ENVR & ENVR \\ 
  31 & CSSC & JSS & SMMR & SMMR \\ 
  32 & JBS & SMMR & MTKA & MTKA \\ 
  33 & JTSA & JBS & CSSC & CSSC \\ 
  34 & JAS & JTSA & JTSA & JSS \\ 
  35 & JSS & MTKA & JBS & JTSA \\ 
  36 & MTKA & ISR & STATS & ANZS \\ 
  37 & SMMR & CSSC & JSS & JBS \\ 
  38 & ISR & ANZS & ANZS & STATS \\ 
  39 & STATS & STATS & ISR & ISR \\ 
  40 & ANZS & JAS & JAS & JAS \\ 
  41 & JABES & JABES & JABES & CMPST \\ 
  42 & EES & EES & CMPST & JABES \\ 
  43 & STATAJ & CMPST & STMOD & STMOD \\ 
  44 & CMPST & STMOD & STNEE & STNEE \\ 
  45 & STMOD & STNEE & EES & EES \\ 
  46 & STPAP & STATAJ & STPAP & STPAP \\ 
  47 & STNEE & STPAP & STATAJ & STATAJ \\ 
   \hline
\end{tabular}
\caption{\footnotesize{Total influence ranking. PSJR: Prestige Scimago Journal Rank (self-citations restricted to 33\% of references); EBPR: Empirical Bayes PageRank (self-citations included);
EIFA*: Eigenfactor; EBEF*: Empirical Bayes Eigen Factor
* without self-citations. Journals are ordered according to EBEF scores. }} \label{fig:ranktotalR}
\end{table}

\begin{table}[ht]
\centering
\scriptsize
\begin{tabular}{rllll}
  \hline
 & EBPR & PSJR & EIFA & EBEF \\ 
  \hline
1 & JRSS-B & JRSS-B & JRSS-B & JRSS-B \\ 
  2 & AOS & AOS & STSCI & STSCI \\ 
  3 & STSCI & STSCI & AOS & AOS \\ 
  4 & JASA & JASA & JASA & JASA \\ 
  5 & BKA & BKA & BKA & BKA \\ 
  6 & BIOST & BIOST & BIOST & BIOST \\ 
  7 & BCS & BCS & SJS & SJS \\ 
  8 & JCGS & JCGS & BCS & JCGS \\ 
  9 & SJS & SJS & JCGS & BCS \\ 
  10 & STSIN & STSIN & STSIN & STSIN \\ 
  11 & TEST & TEST & TEST & TEST \\ 
  12 & CJS & CJS & CJS & CJS \\ 
  13 & STCMP & STCMP & STCMP & STCMP \\ 
  14 & TECH & BERN & BERN & BERN \\ 
  15 & BERN & TECH & LTA & TECH \\ 
  16 & JRSS-A & JRSS-A & TECH & LTA \\ 
  17 & LTA & LTA & JRSS-A & JRSS-C \\ 
  18 & JMA & JMA & JMA & JRSS-A \\ 
  19 & STMED & JRSS-C & JRSS-C & JMA \\ 
  20 & JRSS-C & STMED & SMMR & STMOD \\ 
  21 & AMS & AMS & STMOD & SMMR \\ 
  22 & ISR & ISR & AISM & AMS \\ 
  23 & SMMR & SMMR & AMS & ISR \\ 
  24 & JSS & STMOD & ISR & AISM \\ 
  25 & CSDA & JSS & STMED & JNS \\ 
  26 & AISM & CSDA & JNS & STMED \\ 
  27 & STMOD & AISM & BIOJ & BIOJ \\ 
  28 & JNS & JNS & JSPI & CSDA \\ 
  29 & BIOJ & BIOJ & CSDA & JSS \\ 
  30 & JSPI & JSPI & JSS & JSPI \\ 
  31 & EES & EES & JABES & JABES \\ 
  32 & JTSA & JABES & ANZS & ANZS \\ 
  33 & JABES & JTSA & MTKA & STNEE \\ 
  34 & MTKA & ANZS & JTSA & STATS \\ 
  35 & ENVR & ENVR & STATS & EES \\ 
  36 & STATAJ & MTKA & EES & MTKA \\ 
  37 & ANZS & STNEE & STNEE & JTSA \\ 
  38 & STATS & STATS & ENVR & ENVR \\ 
  39 & STNEE & CMPST & CMPST & CMPST \\ 
  40 & JBS & JBS & JSCS & JSCS \\ 
  41 & SPL & SPL & SPL & SPL \\ 
  42 & CMPST & JSCS & JBS & JBS \\ 
  43 & JSCS & STATAJ & CSSC & CSSC \\ 
  44 & CSTM & CSTM & CSTM & CSTM \\ 
  45 & CSSC & CSSC & STATAJ & STATAJ \\ 
  46 & STPAP & STPAP & STPAP & STPAP \\ 
  47 & JAS & JAS & JAS & JAS \\ 
   \hline
\end{tabular}
\caption{\footnotesize{Articles-level influence ranking. PSJR: Prestige Scimago Journal Rank (self-citations restricted to 33\% of references); EBPR: Empirical Bayes PageRank (self-citations included);
EIFA*: Eigenfactor; EBEF*: Empirical Bayes Eigen Factor
* without self-citations. Journals are ordered according to EBEF scores.} } \label{fig:rankarticleR}
\end{table}

\subsection{Half sampling procedure} \label{app:half}

We generate training matrices 
according to a Beta-Bernoulli
process such that $\tilde c_{ij}=\sum_{k=1}^{c_{ij}} X_k$, with $X_k \in \{0,1\}$, $P(X_k=0)=q$
and
$q \sim \mbox{Beta}(a,b)$ . Then
$
\E(X_k) = q = \frac{a}{a +b} , \mbox{Var} (X_k) = q(1- q)$ and $\mbox{Cor}(X_k,X_\ell)=\rho=(a+b+1)^{-1}$.
The marginal distribution of $\tilde c_{ij}$ is a Beta-Binomial distribution with parameters $(a,b,c_{ij})$ and has
expectation $\E(\tilde c_{ij})= c_{ij}q$ and $\mbox{Var} (\tilde c_{ij})= c_{ij} q(1- q) [1+ (c -1)\rho]$.
This sampling procedure is similar to the Binomial sampling described in the Discussion but it takes into account the
overdispersion due to an intra-class correlation $\rho$ among binary draws intra
cells. For the  Monte Carlo half sampling we take: $a = b =10$ resulting in $q = 0.5$
and $\rho \approx  0.05$. Other values would have been envisioned, but this one
corresponds to the estimation of an average intra-class correlation
$\mbox{Cor}(X_{ijk}, X_{ij\ell})$ within citing journals $i$ for each category of response $j$ (journal
cited) as defined by \citep{Landis} and estimated via MANOVA
procedures.
%to increase the variability of the subsamples. 

 The two sampling procedures ($\rho=0$ and $\rho=0.05$)
 gave almost the same results both for parameter ML estimation and for influence scores given in Table \ref{fig:half}. 
 Moreover, the results obtained with MC half samples are similar to the results obtained on the complete data set.
 The coefficients $\gamma$ are stable. Nevertheless, we observe that the coefficients $\alpha_i$ are smaller (0.90 for CSDA instead of 0.95 and 0.23 for STATAJ instead of 0.39) which implies a more aggressive shrinkage. This behavior is expected as there are less data available to rely on.

\begin{table}[ht]
\centering
\scriptsize
\begin{tabular}{rrllll}
  \hline
& Journal & EBPR & PSJR & EIFA & EBEF \\ 
  \hline
1 & JRSS-B & 5.29 & 5.77 & 5.84 & 5.82 \\ 
  2 & STSCI & 3.45 & 3.83 & 4.22 & 4.22 \\ 
  3 & AOS & 4.11 & 4.40 & 3.82 & 3.67 \\ 
  4 & JASA & 3.43 & 3.73 & 3.65 & 3.57 \\ 
  5 & BKA & 2.66 & 2.88 & 3.07 & 3.05 \\ 
  6 & BIOST & 1.70 & 1.79 & 1.94 & 1.90 \\ 
  7 & SJS & 1.40 & 1.51 & 1.74 & 1.79 \\ 
  8 & JCGS & 1.51 & 1.61 & 1.70 & 1.73 \\ 
  9 & BCS & 1.70 & 1.78 & 1.73 & 1.72 \\ 
  10 & STSIN & 1.27 & 1.35 & 1.53 & 1.55 \\ 
  11 & TEST & 1.15 & 1.25 & 1.43 & 1.48 \\ 
  12 & CJS & 1.15 & 1.25 & 1.36 & 1.46 \\ 
  13 & STCMP & 1.11 & 1.17 & 1.28 & 1.29 \\ 
  14 & BERN & 1.05 & 1.03 & 1.19 & 1.14 \\ 
  15 & TECH & 1.05 & 1.06 & 1.04 & 1.11 \\ 
  16 & LTA & 0.94 & 0.95 & 1.05 & 1.06 \\ 
  17 & JRSS-C & 0.79 & 0.83 & 0.90 & 0.95 \\ 
  18 & JRSS-A & 1.03 & 0.95 & 0.91 & 0.92 \\ 
  19 & JMA & 0.89 & 0.87 & 0.91 & 0.90 \\ 
  20 & STMOD & 0.64 & 0.71 & 0.71 & 0.85 \\ 
  21 & SMMR & 0.70 & 0.69 & 0.76 & 0.77 \\ 
  22 & AMS & 0.76 & 0.72 & 0.68 & 0.73 \\ 
  23 & ISR & 0.75 & 0.69 & 0.68 & 0.72 \\ 
  24 & AISM & 0.65 & 0.60 & 0.69 & 0.68 \\ 
  25 & JNS & 0.59 & 0.56 & 0.65 & 0.66 \\ 
  26 & STMED & 0.85 & 0.80 & 0.66 & 0.63 \\ 
  27 & BIOJ & 0.58 & 0.56 & 0.59 & 0.61 \\ 
  28 & CSDA & 0.66 & 0.62 & 0.55 & 0.57 \\ 
  29 & JSS & 0.66 & 0.62 & 0.52 & 0.56 \\ 
  30 & JSPI & 0.58 & 0.52 & 0.56 & 0.55 \\ 
  31 & ANZS & 0.46 & 0.44 & 0.47 & 0.52 \\ 
  32 & JABES & 0.50 & 0.47 & 0.48 & 0.52 \\ 
  33 & STATS & 0.44 & 0.40 & 0.46 & 0.48 \\ 
  34 & STNEE & 0.41 & 0.40 & 0.41 & 0.48 \\ 
  35 & MTKA & 0.48 & 0.41 & 0.47 & 0.47 \\ 
  36 & JTSA & 0.54 & 0.43 & 0.47 & 0.46 \\ 
  37 & EES & 0.55 & 0.47 & 0.42 & 0.45 \\ 
  38 & ENVR & 0.46 & 0.41 & 0.41 & 0.44 \\ 
  39 & CMPST & 0.37 & 0.34 & 0.37 & 0.41 \\ 
  40 & JSCS & 0.36 & 0.31 & 0.33 & 0.35 \\ 
  41 & SPL & 0.39 & 0.30 & 0.31 & 0.30 \\ 
  42 & JBS & 0.40 & 0.31 & 0.30 & 0.29 \\ 
  43 & CSSC & 0.28 & 0.20 & 0.23 & 0.24 \\ 
  44 & STPAP & 0.26 & 0.18 & 0.20 & 0.22 \\ 
  45 & CSTM & 0.30 & 0.20 & 0.23 & 0.21 \\ 
  46 & JAS & 0.25 & 0.17 & 0.18 & 0.20 \\ 
  47 & STATAJ & 0.47 & 0.22 & 0.23 & 0.19 \\ 
   \hline
\end{tabular}
\caption{\footnotesize{Articles-level influence score with half sampling procedure. PSJR: Prestige Scimago Journal Rank (self-citations restricted to 33\% of references); EBPR: Empirical Bayes PageRank (self-citations included);
EIFA*: Eigenfactor; EBEF*: Empirical Bayes Eigen Factor
* without self-citations. Journals are ordered according to EBEF scores.}} \label{fig:half}
\end{table}

\section*{Acknowledgements}
Thanks are expressed to Prof John James (University of Sydney)  for his comments on  an earlier version of the manuscript. 

\bibliographystyle{Chicago}
\bibliography{references}
\end{document}